\documentclass[epsfig]{article}
\usepackage{amsfonts}

\usepackage{graphicx}
\usepackage{amsmath}

\input epsf.sty
\newtheorem{theorem}{Theorem}
\newtheorem{acknowledgement}[theorem]{Acknowledgement}

\input{tcilatex}

\begin{document}

\title{\rightline{\mbox {\normalsize
{Lab/UFR-HEP0308/GNPHE/0309/IFT-UAM/CSIC-03-28}}} \textbf{Classification of }%
$\mathcal{N}\mathbf{=2}$\textbf{\ supersymmetric CFT}$_{4}$\textbf{s}:\\
\textbf{Indefinite Series}}
\author{M. Ait Ben Haddou$^{1,2,3}$\thanks{%
aitbenha@fsmek.ac.ma}, A. Belhaj$^{4}$\thanks{%
ufrhep@fsr.ac.ma} \& E.H. Saidi$^{1,3}$\thanks{%
h-saidi@fsr.ac.ma} \\
{\small 1 Lab/UFR-Physique des Hautes Energies, Facult\'{e} Sciences de
Rabat, Morocco.}\\
{\small 2} {\small D\'{e}partement de Math\'{e}matique \& Informatique,}
{\small Fac Sciences de Meknes, Morocco,}\\
{\small 3} {\small Groupement National de Physique Hautes Energies, GNPHE;
si\`{e}ge focal, Fac Sc Rabat, Morocco.}\\
{\small 4-Instituto de Fisica Teorica, C-XVI, Universidad Autonoma de
Madrid, E-28049-Madrid, Spain.}}
\maketitle

\begin{abstract}
Using geometric engineering method of $4D$ $\mathcal{N}=2$ quiver gauge
theories and results on the classification of Kac-Moody (KM) algebras, we
show on explicit examples that there exist three sectors of $\mathcal{N}=2$
infrared CFT$_{4}$s. Since the geometric engineering of these CFT$_{4}$s
involve type II strings on K3 fibered CY3 singularities, we conjecture the
existence of three kinds of singular complex surfaces containing, in
addition to the two standard classes, a third indefinite set. To illustrate
this hypothesis, we give explicit examples of K3 surfaces with H$_{3}^{4}$
and E$_{10}$ hyperbolic singularities. We also derive a hierarchy of
indefinite complex algebraic geometries based on affine $A_{r}$ and T$%
_{\left( p,q,r\right) }$ algebras going beyond the hyperbolic subset. Such
hierarchical surfaces have a remarkable signature that is manifested by the
presence of poles.\newline

\end{abstract}

\thispagestyle{empty} \newpage \setcounter{page}{1}

\newpage

\section{Introduction}

Recently $D$ dimension supersymmetric conformal field theories (CFT$_{D}$ )
have been subject to an intensive interest in connection with superstring
compactifications on Calabi-Yau (CY) manifolds \cite{1}-\cite{4} and AdS/CFT
correspondence \cite{5,6}. An important class of these super CFTs
corresponds to those embedded in type II string compactifications on K3
fibered CY threefolds (CY3) with $ADE$ singularities. These theories admit a
very nice geometric engineering \cite{7,8} in terms of quiver diagrams and
are classified into two categories according to the type of K3
singularities: (\textbf{a}) $\mathcal{N}=2$ CFT$_{4}$s with gauge group $%
G=\prod_{i}SU\left( s_{i}n\right) $ and bi-fundamental matters. This
category of scale invariant field models is classified by \textit{affine} $%
\widehat{ADE}$ Lie algebras. They have vanishing individual beta function $%
b_{i}$ known to be given by $b_{i}=\frac{1}{12}\left( 44n_{i}-\sum_{j}\left[
8a_{ij}^{4}+2a_{ij}^{6}\right] n_{j}\right) $ with $a_{ij}^{4}$ and $%
a_{ij}^{6}$\ being the number of Weyl fermions and scalars respectively \cite
{2,9}. In $\mathcal{N}=2$ affine CFT$_{4}$s, this beta function relation can
be put in the form\ $b_{i}=\frac{11}{6}\mathbb{K}_{ij}^{\left( 0\right)
}n_{j}$\ and its vanishing condition $\mathbb{K}_{ij}n_{j}=0$ can be solved
in terms of the usual Dynkin integer weights $s_{i}$ ($\mathbb{K}%
_{ij}s_{j}=0 $) as follows,
\begin{equation}
\mathbb{K}_{ij}^{\left( 0\right) }n_{j}=n\mathbb{K}_{ij}^{\left( 0\right)
}s_{j}=0,  \label{1}
\end{equation}
where $\mathbb{K}_{ij}^{\left( 0\right) }$ is the affine $\widehat{ADE}$
Cartan matrix. The extra upper index on $\mathbb{K}_{ij}^{\left( 0\right) }$
is introduced for later use. (\textbf{b}) $\mathcal{N}=2$ CFT$_{4}$s, based
on \textit{finite} $ADE$ singularities; with gauge group $%
G=\prod_{i}SU\left( n_{i}\right) $ and matters in both fundamental $\mathbf{n%
}_{i}$ and bi-fundamental $\left( \mathbf{n}_{i}\mathbf{,}\overline{\mathbf{n%
}}_{j}\right) $ representations of $G$. In this case, the beta function $%
b_{i}$ may be put in the form $b_{i}=\frac{11}{6}\left( \mathbb{K}%
_{ij}^{\left( +\right) }n_{j}-m_{i}\right) $ and so its vanishing condition
is equivalent to,
\begin{equation}
\mathbb{K}_{ij}^{\left( +\right) }n_{j}=+m_{i},  \label{2}
\end{equation}
where now $\mathbb{K}_{ij}^{\left( +\right) }$ is the finite $ADE$ Cartan
matrix and where $m_{i}$ is interpreted as the number of fundamental
matters. Here also, we have introduced the extra upper index on $\mathbb{K}%
_{ij}^{\left( +\right) }$ to distinguish it from $\mathbb{K}_{ij}^{\left(
0\right) }$ of eq(\ref{1}). Note that eq(\ref{2}) may be thought of as a
special deformation of eq(\ref{1}), which in field theoretic language,
consists to \textit{add} a definite number of Weyl fermions and scalars;
that is more supersymmetric fundamental matters. This interpretation is not
a new idea in QFT$_{d}$; something close to that was already used in the
study of deformations of the $2D$ conformal structure; in particular in the
analysis of deformations of 2D Toda field theories. In the present $4D$
case, much informations on the deformation of eq(\ref{1}) to eq(\ref{2}) and
vice versa may be read directly on the explicit relation $b_{i}=\frac{1}{12}%
\left( 44n_{i}-\vartheta _{i}\right) $ with $\vartheta _{i}=\sum_{j}\left[
8a_{ij}^{4}+2a_{ij}^{6}\right] n_{j}$. Starting from $b_{i}>0$; that is $%
44n_{i}>\vartheta _{i}$, one can recover conformal invariance by adding
appropriate amount of fundamental matter to the quiver gauge system; this
corresponds to increasing $\vartheta _{i}$ until to reach the conformal
point. Pushing this reasoning further by remarking that as one may \textit{%
add} matter, one may also \textit{integrate it out}. This corresponds to
starting from $b_{i}<0$, i.e $44n_{i}<\vartheta _{i}$ and integrating out
some amount of matter which discreases\ $\vartheta _{i}$. The resulting beta
function can be put in the form $\frac{11}{6}\left( \mathbb{K}_{ij}^{\left(
-\right) }n_{j}+m_{i}\right) $; so one ends with the following conformal
invariant dual formula to eq(\ref{2}),
\begin{equation}
\mathbb{K}_{ij}^{\left( -\right) }n_{j}=-m_{i};\qquad i=1,....  \label{a3}
\end{equation}
To give an interpretation to $\mathbb{K}_{ij}^{\left( -\right) }$ matrix,
note that the above three eqs show that they are really very remarkable
relations in the sense that they may be put altogether into a condensed form
as follows
\begin{equation}
\mathbb{K}_{ij}^{\left( q\right) }n_{j}=qm_{i};\qquad q=+1,0,-1.  \label{a4}
\end{equation}
But this formula is very well known in the literature on KM algebras as it
is just the statement of the theorem of their classification which says that
the three $q=+1,0,-1$ sectors correspond respectively to finite, affine and
indefinite classes of KM algebras \cite{10}.

In this paper, we develop the study for the particular class of indefinite $%
\mathcal{N}=2$ CFT$_{4}$s. We will show that this class shares all the basic
features we know about finite and affine $\mathcal{N}=2$ QFT$_{4}$s and
their IR CFT$_{4}$ limits embedded in type II string on CY3 with singular K3
fibration. As a consequence of this classification, we conjecture the
existence of a third class of local K3s with indefinite singularities; the
two others are the known $ADE$ ones. As we usually do in finite and affine
standard cases, we will focus our attention here also on the simply laced
subset of local K3s classified by indefinite KM algebras and the
corresponding mirror geometries. More precisely, we study the special case
of $\mathcal{N}=2$ CFT$_{4}$ models based on simply laced hyperbolic
symmetries as well as particular extensions.

The presentation of this paper is as follows: In section $2$, we review
briefly the computation of the general expression of beta function of $%
\mathcal{N}=2$ QFT$_{4}$s using geometric engineering method. Then, we show
that the solution for $\mathcal{N}=2$ CFT$_{4}$ scale invariance condition
coincides exactly with the Lie algebraic classification eq(\ref{a4}). In
sections $3$ and $4$, we establish a classification theorem for $\mathcal{N}%
=2$ CFT$_{4}$s and give two explicit illustrating examples. These concern
local K3 with hyperbolic H$_{3}^{4}$ and E$_{10}$ singularities. In section $%
5$, we give a conclusion and generalizations.

\section{Beta Function in $\mathcal{N}=2$ quiver QFT$_{4}$}

A nice way to compute the beta function of the $\mathcal{N}=2$ quiver gauge
theories is to use the geometric engineering method of QFT$_{4}$s embedded
in type II strings on CY3 with $ADE$ singularities \cite{7}. This method
involves toric representation of CY3, mirror symmetry and techniques of
algebraic geometry; in particular trivalent geometry which we review its
main lines here below. Details can be found in \cite{7,8}. To illustrate the
idea of the method in a comprehensive way, we start by considering the case
of a unique trivalent vertex; then we give the results for chains of
trivalent vertices.

\textbf{Case of one trivalent vertex: \qquad }In type IIA string on CY3, a
typical trivalent vertex of the toric representation of CY3 is described by
the 3-dimensional vectors $V_{i}$,
\begin{equation}
V_{0}=\left( 0,0,0\right) ;\quad V_{1}=\left( 1,0,0\right) ;\quad
V_{2}=\left( 0,1,0\right) ;\quad V_{3}=\left( 0,0,1\right) ;\quad
V_{4}=\left( 1,1,1\right)  \label{4}
\end{equation}
satisfying the following toric geometry relation $%
\sum_{i=0}^{4}q_{i}V_{i}=-2V_{0}+V_{1}+V_{2}+V_{3}-V_{4}=0$. The vector
charge $\left( q_{i}\right) =\left( -2,1,1,1,-1\right) $ is known as the
Mori vector and the sum of its $q_{i}$ components is zero as required by the
CY condition. In type IIB mirror geometry, the $\left(
V_{0},V_{1},V_{2},V_{3},V_{4}\right) $ vertices are\ represented by complex
variables $\left( u_{0},u_{1},u_{2},u_{3},u_{4}\right) $ constrained as $%
\prod_{i}u_{i}^{q_{i}}=1$ and solved by $\left( 1,x,y,z,xyz\right) $; see
figure 1.
\begin{figure}[tbh]
\begin{center}
\epsfxsize=4cm \epsffile{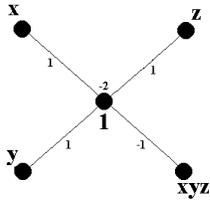}
\end{center}
\caption{{\protect\small \textit{This is a typical trivalent vertex in
mirror geometry; It involves a central node and 4 attached ones; 2 of them
are of Dynkin type and the others are required by CY \ condition. They deal
with fundamental matters.}}}
\end{figure}
In terms of these variables, the algebraic geometry eq describing mirror
geometry is given by the following complex surface, $P\left( X^{\ast
}\right) =\mathrm{e}_{0}+\mathrm{a}_{0}x+\mathrm{b}_{0}y+\left( \mathrm{c}%
_{0}-\mathrm{d}_{0}xy\right) z$, where $\mathrm{a}_{0}\mathrm{,b}_{0}\mathrm{%
,c}_{0}\mathrm{,d}_{0}$ and $\mathrm{e}_{0}$ are non zero complex moduli.
Upon eliminating the $z$ variable by using the eq of motion $\frac{\partial
P\left( X^{\ast }\right) }{\partial z}=0$, the above trivalent geometry
reduces exactly to
\begin{equation}
P\left( X^{\ast }\right) =\mathrm{a}_{0}x+\mathrm{e}_{0}+\frac{\mathrm{b}_{0}%
\mathrm{c}_{0}}{\mathrm{d}_{0}}\frac{1}{x},  \label{7}
\end{equation}
which is nothing but the mirror of the su$\left( 2\right) $ singularity of
local K3 surface. To get the eq of the CY3, one promotes the coefficients $%
\mathrm{a}_{0}\mathrm{,b}_{0}\mathrm{,c}_{0}\mathrm{,d}_{0}$ and $\mathrm{e}%
_{0}$ to holomorphic polynomials on complex plane as,

\begin{equation}
\mathrm{e}=\sum_{i=0}^{n_{r}}\mathrm{e}_{i}\zeta ^{i};\quad \mathrm{a}%
=\sum_{i=0}^{n_{r-1}}\mathrm{a}_{i}\zeta ^{i};\quad \mathrm{b}%
=\sum_{i=0}^{n_{r+1}}b_{i}\zeta ^{i},\quad \mathrm{c}%
=\sum_{i=0}^{m_{r}}c_{i}\zeta ^{i};\quad \mathrm{d}=\sum_{i=0}^{m_{r}^{%
\prime }}d_{i}\zeta ^{i}.  \label{8}
\end{equation}
Note that the functions $\mathrm{a,b}$ and $\mathrm{e}$ encode the
fibrations of $SU\left( 1+n_{r-1}\right) \times SU\left( 1+n_{r}\right)
\times SU\left( 1+n_{r+1}\right) $ gauge symmetry while $\mathrm{c}$ and $%
\mathrm{d}$ are associated with flavor symmetries of the underlying $%
\mathcal{N}=2$ QFT$_{4}$ engineered over the nodes of the trivalent vertex.
The nature of the flavor group will be discussed later on; all what we know
about it is that for $m_{r}^{\prime }=0$, the group is $SU\left(
1+m_{r}\right) $ but this corresponds to finite class of $\mathcal{N}=2$ CFT$%
_{4}$s. Note also that in geometric engineering method, the $SU\left(
1+n_{r}\right) $ and $SU\left( 1+n_{r\pm 1}\right) $ gauge symmetries are
fibered over $V_{0}$, $V_{1}$ and $V_{2}$. However the two kind of
''matters'' $m_{r}$ and $m_{r}^{\prime }$ are fibered over the nodes $V_{3}$
and $V_{4}$ respectively, see figure2.\bigskip
\begin{figure}[tbh]
\begin{center}
\epsfxsize=8cm \epsffile{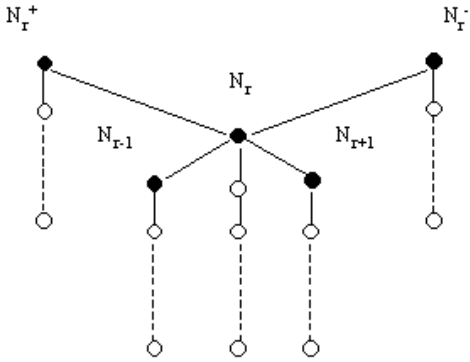}
\end{center}
\caption{{\protect\small \textit{This graph describes a typical vertex one
has in geometric engineering of }}$\mathcal{N}\mathit{=2}$\textit{\ }%
{\protect\small \textit{QFT}}$_{4}$. $SU\left( 1+l\right) ${\protect\small
\textit{\ gauge and flavor symmetries are fibered over the five black nodes.
Flavor symmetries require large base volume.}}}
\end{figure}
Note finally that the holomorphic functions $\mathrm{a,b,c,d}$ and $\mathrm{e%
}$ are not all of them independent, one can usually fix one of them. We will
see that this freedom turns into a condition on $m_{r}$ and $m_{r}^{\prime }$%
; but for the moment, we keep all these moduli free and make a comment later
on.

\textbf{Infrared} $\mathcal{N}=2$ \textbf{QFT}$_{4}$\textbf{\ limit: \qquad }%
To get the various $\mathcal{N}=2$ CFT$_{4}$s embedded in type IIA strings
on CY3, we have to study the infrared field theory limit one gets from
mirror geometry eq(\ref{7}) and look for the scaling properties of the gauge
coupling constant moduli. We will do this explicitly for the case of the
trivalent vertex and then give the general result for the chain. To that
purpose, we proceed in three steps: First determine the behaviour of the
complex moduli $\mathrm{f}_{i}$ appearing in the expansion eq(\ref{8}) under
a shift of $\zeta $ by $1/\varepsilon $ with $\varepsilon \rightarrow 0$.
Doing this and requiring that eqs(\ref{8}) should be preserved, that is
still staying in the singularity described by eqs(\ref{8}), we get the
following,

\begin{equation}
\mathrm{e}_{l}\sim \varepsilon ^{l-n_{r}};\qquad \mathrm{a}_{l}\sim
\varepsilon ^{l-n_{r-1}};\qquad \mathrm{b}_{l}\sim \varepsilon
^{l-n_{r+1}};\qquad \mathrm{c}_{l}\sim \varepsilon ^{l-m_{r}};\qquad \mathrm{%
d}_{l}\sim \varepsilon ^{l-m_{r}^{\prime }}.  \label{9}
\end{equation}
Second compute the scaling behaviour of the\ gauge coupling constant moduli $%
Z^{\left( g\right) }$ under the shift $\zeta ^{\prime }=\zeta +1/\varepsilon
$. Putting eqs(\ref{9}) back into the explicit expression of $Z^{\left(
g\right) }$ namely $Z^{\left( g\right) }=\frac{a_{0}b_{0}c_{0}}{%
e_{0}^{2}d_{0}}$, we get the following behaviour $Z^{\left( g_{r}\right)
}\sim \varepsilon ^{-b_{r}}$ with $b_{r}$ given by,
\begin{equation}
b_{r}=\frac{11}{6}\left[ 2n_{r}-n_{r-1}-n_{r+1}-\left( m_{r}-m_{r}^{\prime
}\right) \right] .  \label{11}
\end{equation}
This relation tells us: (\textbf{i}) $b_{r}$ is the beta function for the
gauge group factor $SU\left( 1+n_{r}\right) $. (\textbf{ii}) $b_{r}$ depends
on $m_{i}^{\ast }=m_{r}-m_{r}^{\prime }$; it is invariant under global
shifts of $m_{r}$ and $m_{r}^{\prime }$, a property which reflects the
arbitrariness we have referred to above. Introducing the following notation
\textit{sing}$\left( m_{i}^{\ast }\right) =q$ with $q=+1,0,-1$ respectively
associated with the intervals $m_{r}>m_{r}^{\prime }$, $m_{r}=m_{r}^{\prime
} $ and \ $m_{r}<m_{r}^{\prime }$, we can rewrite eq(\ref{11}) as $\mathbf{K}%
_{ij}^{\left( q\right) }n_{j}-q\left| m_{i}^{\ast }\right| $; see also eq(%
\ref{a4}). Finally taking the limit $\varepsilon \rightarrow 0$, finiteness
of $Z^{\left( g\right) }$ requires then that the field theory limit should
be asymptotically free; that is $b_{r}\leq 0$. Upper bound $b_{r}=0$
corresponds to scale invariance we are interested in here.

\textbf{Conformal Invariance phases: \qquad }From eq(\ref{11}) it is not
difficult to recognize the three classes of solutions for $\mathbf{K}%
_{ij}^{\left( q\right) }n_{j}=qm_{i}^{\ast }$: \textbf{(i)} $%
m_{r}-m_{r}^{\prime }=0$ and $n_{r}=n_{r-1}=n_{r+1}=n$; this corresponds to
a generic vertex of $\widehat{SU\left( k\right) }$ affine $\mathcal{N}=2$\
conformal CFT$_{4}$ with $SU\left( n\right) ^{3}$ gauge symmetry. Extension
to the other $\widehat{DE}$ geometries is straightforward. \textbf{(ii)} $%
m_{r}^{\prime }=0$, but the other integers may be taken as $n_{r}=\alpha n;$
$n_{r-1}=\beta n,\ n_{r+1}=\gamma n,$ $m_{r}=\delta n$ with $\alpha ,$ $%
\beta $, $\gamma $, $\delta $ $\in n\mathbb{Z}_{+}$ constrained as $2\alpha
=\beta +\gamma +\delta $. As an example, one may take them as $%
m_{r}=n_{r-1}=n_{r+1}=2n$ and $n_{r}=3n$; this corresponds to a gauge
symmetry $SU\left( 3n\right) \times SU\left( 2n\right) ^{2}$ and an $%
SU\left( 2n\right) $ flavor symmetry engineered on the middle vertex of the
SU$\left( 4\right) $ finite Dynkin diagrams. This solution is also valid for
$m_{r}-m_{r}^{\prime }>0$; all one has to do is to substitute the expression
of $m_{r}$ of the above solution by $m_{r}^{\ast }$. \textbf{(iii)} For the
remarkable case $m_{r}=0$; that is $m_{r}^{\ast }<0$,\ conformal invariance
requires $2n_{r}-n_{r-1}-n_{r+1}+m_{r}^{\prime }=0$ and is solved as $%
n_{r}=\alpha n;$ $n_{r-1}=\beta n,\ n_{r+1}=\gamma n,$ $m_{r}^{\prime
}=\delta ^{\prime }n$ with $\alpha ,$ $\beta $, $\gamma $, $\delta ^{\prime
} $ $\in n\mathbb{Z}_{+}$ satisfying $2\alpha +\delta ^{\prime }=\beta
+\gamma $. As an example, one may take them as $m_{r}^{\prime
}=n_{r-1}=n_{r+1}=2n$ and $n_{r}=n$. Note that solutions for conformal
invariance may have $m_{r}^{\prime }>n_{r}$ as one sees on the above
particular solution. This property constitute one of the arguments we will
use to conjecture the flavor symmetry $SU\left( qm_{r}^{\ast }\right) $; it
recovers the known results as particular cases. Naturally the $q=-1$ sector
corresponds to a new class of solutions. In this regards we will show that
this class is linked with simply laced indefinite KM algebras. To do so we
need however more than one trivalent vertex since simply laced indefinite
Lie algebras have at least a rank four and this corresponds to the over
extension of affine $\widehat{A}_{2}$.

\textbf{Chains of trivalent vertices: \qquad }To get the generalization of
the above results, it is enough to think about the previous vertices as a
generic trivalent vertex of a linear chain of $N$ trivalent vertices, that
is
\begin{equation}
V_{0}\rightarrow V_{\alpha }^{0};\qquad V_{3}\rightarrow V_{\alpha
}^{+};\qquad V_{4}\rightarrow V_{\alpha }^{-};\qquad V_{1}\rightarrow
V_{\alpha -1}^{0};\qquad V_{2}^{0}\rightarrow V_{\alpha +1}^{0},  \label{12}
\end{equation}
where $\alpha \in \left\{ 1,...,N\right\} $. The intersections between $%
V_{\alpha }^{0}$ and $V_{\alpha \pm 1}^{0}$ are specified by some integers $%
\mathrm{q}_{\alpha }^{i}$ generally inspired from the Cartan matrix of the
KM algebra one is interested in. In this generic case, the data of the toric
polytope are fixed by $\sum_{\alpha \geq 0}\left( \mathrm{q}_{\alpha
}^{i}V_{\alpha }^{0}+V_{i}^{+}-V_{i}^{-}\right) =0$ and $\sum_{\alpha }^{i}%
\mathrm{q}_{\alpha }^{i}=0$. Note that the $\pm $ upper indices carried by
the $V_{i}^{\pm }$ vertices refer to the fourth $+1$ and five $-1$ entries
of the Mori vector $\mathrm{q}_{\tau }^{i}=\left( \mathrm{q}_{\alpha
}^{i};+1,-1\right) $ of trivalent vertex. In practice, the Mori vectors $%
\mathrm{q}_{\alpha }^{i}$s form a $N\times \left( N+s\right) $ rectangular
matrix whose $N\times N$ square sub-matrix $q_{j}^{i}$ is minus the
generalized Cartan matrix $\mathbb{K}_{ij}^{\left( q\right) }$. For the
example of affine $A_{N-1}$, the Mori charges read as $\mathrm{q}_{\alpha
}^{i}=2\delta _{\alpha }^{i}-\delta _{\alpha }^{i-1}-\delta _{\alpha }^{i+1}$
with the usual periodicity of affine $\widehat{SU\left( N\right) }$. The
remaining $N\times s$ part of $\mathrm{q}_{\alpha }^{i}$ is fixed by the CY
condition $\sum_{\alpha }^{i}\mathrm{q}_{\alpha }=0$ and the corresponding
vertices are interpreted as dealing with non compact two dimension divisors
defining the singular space on which live singularities. In mirror geometry
where $x_{\alpha -1},$\ $x_{\alpha }$,$\ x_{\alpha +1},\ y_{\alpha },$\ and $%
\frac{x_{\alpha -1}x_{\alpha +1}y_{\alpha }}{y_{\alpha }^{2}}$\ are the
variables associated with the vertices (\ref{12}), algebraic eq for a
generic vertex extends as \textrm{a}$_{\alpha -1}x_{\alpha -1}+\mathrm{a}%
_{\alpha }x_{\alpha }+$\textrm{a}$_{\alpha +1}x_{\alpha +1}+\mathrm{c}%
_{\alpha }y_{\alpha }+\mathrm{d}_{\alpha }\frac{x_{\alpha -1}x_{\alpha
+1}y_{\alpha }}{x_{\alpha }^{2}}=0$ where $\mathrm{a}_{\alpha }$, $\mathrm{c}%
_{\alpha }$ and $\mathrm{d}_{\alpha }$ are complex moduli. Summing over the
vertices and setting $y_{\alpha }=x_{\alpha }z_{\alpha }$, one gets $P\left(
X^{\ast }\right) =\mathrm{a}_{0}x_{0}+$ $\sum_{\alpha \geq 1}\left( \mathrm{a%
}_{\alpha }x_{\alpha }+\mathrm{c}_{\alpha }x_{\alpha }z_{\alpha }+\mathrm{d}%
_{\alpha }\frac{x_{\alpha -1}x_{\alpha +1}z_{\alpha }}{x_{\alpha }}\right) $%
. Eliminating the variable $z_{\alpha }$ as we have done for eq(\ref{7}), we
obtain
\begin{equation}
P\left( X^{\ast }\right) =\sum_{\alpha \geq 0}x^{\alpha }\mathrm{a}_{\alpha
}\left( w\right) \prod_{\beta \geq 1}\left( \frac{\mathrm{c}_{\beta }\left(
w\right) }{\mathrm{d}_{\beta }\left( w\right) }\right) ^{\alpha -\beta }.
\label{131}
\end{equation}
From this relation, one gets behaviour $Z^{\left( g_{r}\right) }\sim
\varepsilon ^{-b_{r}}$ with $b_{r}$ given by,
\begin{equation}
b_{r}^{\left( q\right) }=\frac{11}{6}\left[ 2n_{r}-n_{r-1}-n_{r+1}-q\left|
m_{r}^{\ast }\right| \right] ;\qquad r=1,.....  \label{a10}
\end{equation}

\section{Classification Theorem of $\mathcal{N}=2$ CFT$_{4}$s}

Let $\mathcal{G}_{q}$ be some given \textit{simply laced }Lie algebra of
rank \textrm{r}$_{q}=rank\left( \mathcal{G}_{q}\right) $ and Cartan matrix $%
\mathbb{K}^{\left( q\right) }$, $corank\left( \mathbb{K}^{\left( q\right)
}\right) \leq 1$ and let $q=+1$, $0$ and $-1$ be an integer which refers
respectively to the three possible sectors of $\mathcal{G}_{q}$ that is
\textit{finite, affine and indefinite} types. Then the previous results on $%
\mathcal{N}=2$ quiver gauge CFT$_{4}$s can be stated as a theorem to which
we shall refer hereafter as the classification theorem of $\mathcal{N}=2$ CFT%
$_{4}$s. As these supersymmetric gauge theories are special limits of
underlying 4D massive field theories (QFT$_{4}$), we will state this theorem
in a more general way.

\textbf{Theorem:}

For any quiver graph $\Delta \left( \mathcal{G}_{q}\right) $ of trivalent
vertices with a topology type Dynkin diagram of the \textit{simply laced} (
finite, affine and indefinite) Lie algebras $\mathcal{G}_{q}$, there
corresponds:

\textbf{(a)} A $\mathcal{N}=2$ quiver gauge QFT$_{4}$s which is built as
usual by extending the geometric engineering method to include indefinite
type Dynkin diagrams. They may be denoted as QFT$_{4}^{\left( q\right) }$

(\textbf{b}) The quiver gauge group of these $\mathcal{N}=2$ QFT$%
_{4}^{\left( q\right) }$s is $\prod_{i=1}^{\mathrm{r}_{q}}SU\left(
n_{i}\right) $ and the flavor symmetry encoding fundamental matters reads as
$\prod_{i=1}^{\mathrm{r}_{q}}SU\left( qm_{i}^{\ast }\right) $. Here, the
positive integer $\left| m_{i}^{\ast }\right| $ is the effective number of
fundamental matter that contribute to the beta function; it depends on the
absolute value of the difference of $m_{i}$ and $m_{i}^{\prime }$.

(\textbf{c}) The $b_{r}$ functions of the $SU\left( n_{i}\right) $ gauge
symmetries of these $\mathcal{N}=2$ quiver QFT$_{4}$s read as,
\begin{equation}
b_{r}^{\left( q\right) }=\frac{11}{6}\left( \mathbb{K}_{rs}^{\left( q\right)
}n_{s}-q\left| m_{r}^{\ast }\right| \right) ,\qquad r=1,2,...,\mathrm{r}_{q}
\label{t1}
\end{equation}
where $q$ refers to the the above three mentioned sectors.

\textbf{(d)} In the infrared limit of $\mathcal{N}=2$ gauge quiver QFT$_{4}$%
s where $b_{r}^{\left( q\right) }$ $\ \longrightarrow $ $\ 0$, these
theories flow to three classes of 4D $\mathcal{N}=2$ quiver conformal field
theories. The flows are in one to one correspondence with the three sectors
of $\mathcal{G}_{q}$s. As such $\mathcal{N}=2$ CFT$_{4}$s are classified as
QFT$_{4}^{\left( q\right) }$:

(\textbf{i}) \textit{Finite} $ADE$ $\mathcal{N}=2$ CFT$_{4}^{+}$s for which
the vanishing of the beta function leads to $\mathbb{K}_{rs}^{\left(
+\right) }n_{s}=\left| m_{r}^{\ast }\right| $.

(\textbf{ii}) \textit{Affine} $ADE$ $\mathcal{N}=2$ quiver CFT$_{4}^{0}$s
governed by $\mathbb{K}_{rs}^{\left( 0\right) }n_{s}=0$ with one dimension
corank$\left( \mathbb{K}_{rs}^{\left( 0\right) }\right) $.

(\textbf{iii}) \textit{Indefinite} $\mathcal{N}=2$ quiver CFT$_{4}^{-}$s.
They are associated with the class $\mathbb{K}_{rs}^{\left( -\right)
}n_{s}=-\left| m_{r}^{\ast }\right| $ where now $\mathbb{K}_{rs}^{\left(
-\right) }$ is an indefinite Cartan matrix.

To prove this theorem, note that the first three properties follow naturally
from the algebraic geometry analysis of the $\mathcal{N}=2$ quiver QFT$_{4}$%
s embedded in type IIA string on CY3 \cite{7,8} and refs therein. The fourth
property (\textbf{d}) of this theorem can be linked to the Vinberg-Kac-Moody
basic theorem on the classification of Lie algebras which we recall here
below. The property (\textbf{d}) follows from it by setting $u_{i}=n_{i}$
and $v_{i}=\left| m_{r}^{\ast }\right| .$

\textbf{Vinberg-Kac-Moody Theorem }\newline
A generalized indecomposable Cartan matrix $\mathbb{K}$ obey one and only
one of the following three statements: (\textbf{i}) \textit{Finite type (}$%
\det \mathbb{K}>0$ ): There exist a real positive definite vector $\mathbf{u}
$ ($u_{i}>0;$ $i=1,2,...$) such that $\mathbb{K}_{ij}u_{j}=v_{j}>0$. (%
\textbf{ii}) \textit{Affine type, }corank$\left( \mathbb{K}\right) =1$, $%
\det \mathbb{K}=0$\textit{: }There exist a unique, up to a multiplicative
factor, positive integer definite vector $\mathbf{u}$ ($u_{i}>0;$ $i=1,2,...$%
) such that $\mathbb{K}_{ij}u_{j}=0$. (\textbf{iii}) \textit{Indefinite type
(}$\det \mathbb{K}\leq 0$), corank$\left( \mathbb{K}\right) \neq 1$\textit{:
}There exist a real positive definite vector $\mathbf{u}$ ($u_{i}>0;$ $%
i=1,2,...$) such that $\mathbb{K}_{ij}u_{j}=-v_{i}<0.$

All the eqs appearing in this theorem combine together to give eq(\ref{a4}).
As a consequence of this classification of $\mathcal{N}=2$ CFT$_{4}^{\left(
q\right) }$s, our theorem may also be viewed as a classification of possible
K3 singularities. We have then the following,

\textbf{Corollary}

From the property (\textbf{d}) of our classification theorem, we conjecture
the existence of indefinite singularities for K3 fibered CY threefolds that
are characterized by simply laced indefinite Lie algebras. With this
hypothesis, we have: ($\alpha $) \textit{Finite} $ADE$ singularities; ($%
\beta $) \textit{Affine} $\widehat{ADE}$ singularities; ($\gamma $) \textit{%
Indefinite} singularities

Note that the above two first singular K3 surfaces are well common in type
II strings on CY3. However the third one is a new class which to our
knowledge was not studied before. It is dictated from $\mathcal{N}=2$ field
theoretic analysis of $\mathcal{N}=2$ CFT$_{4}^{\left( q\right) }$ possible
solutions. In \cite{11}, we have made a general analysis of such kind of
singularities; here we give explicit illustrating examples. They concern the
over extension of affine $\widehat{A}_{2}$ and the over extension of $%
\widehat{E}_{8}$ respectively baptized as $H_{4}^{3}$ and $E_{10}$.

\section{Two Examples of Hyperbolic Singularities}

We begin by recalling that mirror geometry of type IIA string on CY3 ($X_{3})
$ with affine $\widehat{ADE}$ singularities are conveniently described in
algebraic geometry. A typical eq of such geometry is $P\left( X_{3}^{\ast
}\right) =\sum_{\alpha }a_{\alpha }y_{\alpha }$, where $X_{3}^{\ast }$ is
the mirror of $X_{3}\ $\ and where $a_{\alpha }=a_{\alpha }\left( w\right) $
are complex moduli with expansion similar to those of eq(\ref{8}), see also
\cite{7}. In this relation, the $y_{\alpha }$ complex variables are
constrained as,
\begin{equation}
\prod_{j=1}^{n}y_{j}^{\mathrm{q}_{j}^{i}}=\prod_{\alpha =n+1}^{n+4}y_{\alpha
}^{-\mathrm{q}_{\alpha }^{i}},  \label{211}
\end{equation}
where $\mathrm{q}_{j}^{i}$ is minus the Cartan matrix $\mathbb{K}_{ij}$ of
the corresponding Lie algebra and $y_{\alpha }$, with $n<\alpha <n+5$, are
four extra complex variables that\ are just the monomials appearing in the
elliptic curve $\mathrm{E}=y^{2}+x^{3}+z^{6}+\mathrm{\mu }xyz=0$ on which
shrinks the affine ADE singularity. Therefore,we have,
\begin{equation}
y_{n+1}=y^{2},\quad y_{n+2}=x^{3},\quad y_{n+3}=z^{6},\quad y_{n+4}=xyz,
\label{2110}
\end{equation}
where $\left( y,x,z\right) $\ are the homogeneous coordinates of the
weighted projective space $\mathbf{WP}^{2}\left( 3,2,1\right) $. The
remaining $n$ complex variables $y_{i}$ definitive the $\widehat{ADE}$
geometry are also solved in terms of the previous $y,x$ and $z$ variables.
Such solutions depend on the $\mathrm{q}_{j}^{i}$ and $\mathrm{q}_{\alpha
}^{i}$ integer charges forming altogether a $n\times \left( n+4\right) $
rectangular matrix as
\begin{equation}
{\large Q}_{\alpha }^{i}=\left(
\begin{array}{ccccc}
\mathrm{q}_{j}^{i}, & \mathrm{q}_{n+1}^{i}, & \mathrm{q}_{n+2}^{i}, &
\mathrm{q}_{n+3}^{i}, & \mathrm{q}_{n+4}^{i}
\end{array}
\right) .  \label{b1}
\end{equation}
The resulting two dimensions geometry $y^{2}+x^{3}+z^{6}+\mathrm{\mu }%
_{0}xyz+\sum_{i=1}^{n}\mathrm{a}_{i}y_{i}=0$ have been studied extensively
in the literature for both trivalent and affine geometries. But here we are
claiming that such analysis applies as well for the indefinite sector of Lie
algebras and deals with the un-explored class of indefinite CFT$_{4}$s. As
the better thing to justify our claim is to give examples, we will start by
recalling some useful features on affine geometries and then study the
indefinite case. Before note that the parameter $\mathrm{\mu }$ appearing in
the algebraic geometry eq of the elliptic curve \textrm{E}$\left( \mathrm{%
\mu }\right) $ is its complex structure. It is fixed to a constant $\mathrm{%
\mu }_{0}$ in the case of affine $ADE$ geometries; but vary in the case
indefinite singularities we are interested in here. More precisely, we will
see that in the case of simply laced hyperbolic geometries, the parameter $%
\mathrm{\mu }$ has to vary on a complex plane parameterized by $w$; i.e $%
\mathrm{\mu =\mu }\left( w\right) $. Under this variation, the initial curve
\textrm{E}$\left( \mathrm{\mu }_{0}\right) $ is now promoted to a complex
surface \textrm{E}$\left[ \mathrm{\mu }\left( w\right) \right] $ which, by
the way, is nothing but the elliptic fibration of K3 $y^{2}+x^{3}+z^{6}+%
\mathrm{\mu }\left( w\right) xyz=0$. Note that, upon appropriate
redefinition of variables, one may rewrite the above algebraic geometry eq
of the elliptic curve into the following equivalent form,
\begin{equation}
y^{2}+x^{3}+\mathrm{\nu }\left( t\right) z^{\prime 6}+xyz^{\prime }=0
\label{b5}
\end{equation}
where now $z^{\prime }=\mathrm{\mu }\left( w\right) z$ and $\mathrm{\nu }%
\left( t\right) z^{\prime 6}=z^{6}$. For instance, if we take $\mathrm{\nu }%
\left( t\right) =t^{-1}=w^{-6}$, then $z^{\prime }$ should be as $z^{\prime
}=wz$ and so $\mathrm{\mu }\left( w\right) =w$. Having these properties in
mind, we turn now to illustrate the building of affine A$_{2}$ geometry and
its hyperbolic over extension.

\textbf{Affine extension of }$\mathbf{A}_{2}$\textbf{\ geometry: \qquad }In
the special case of affine $A_{2}$\ geometry, like all series of affine
ADEs, one starts from the curve \textrm{E}$_{0}=y^{2}+x^{3}+z^{6}+\mathrm{%
\mu }_{0}xyz=0$ of $\mathbf{WP}^{2}\left( 3,2,1\right) $ with fixed complex
structure and look for algebraic geometry eq of affine $A_{2}$\ geometry
which reads as,
\begin{equation}
\widehat{A}_{2}:y^{2}+x^{3}+z^{6}+\mathrm{\mu }_{0}xyz+\left( \mathrm{b}%
y_{1}+\mathrm{c}y_{2}+\mathrm{d}y_{3}\right) =0.  \label{c1}
\end{equation}
Here $\mathrm{b},$ $\mathrm{c}$ and $\mathrm{d}$ are complex moduli which
once taken simultaneously to zero the affine $A_{2}$\ geometry shrinks to
the elliptic curve. To get the explicit expression of the remaining $y_{i}$
gauge invariants, one has to specify the toric data for the present affine $%
A_{2}$ geometry and too particularly the $\mathrm{q}_{j}^{i}$ and $\mathrm{q}%
_{\alpha }^{i}$ charges appearing in eq(\ref{211}). These read as,

\begin{equation}
{\large Q}\left( \widehat{A}_{2}\right) =\left(
\begin{array}{ccccccc}
-2 & 1 & 1 & 0 & 0 & 1 & -1 \\
1 & -2 & 1 & 2 & 1 & 0 & -3 \\
1 & 1 & -2 & 0 & 1 & 0 & -1
\end{array}
\right) .  \label{c2}
\end{equation}
The simplest solution one gets for the constraint eqs(\ref{211} regarding $%
y_{1}$, $y_{2}$ and $y_{3}$ is $y_{1}=z^{3},$ $y_{2}=xz$ and $\ y_{2}=y$.
However this is not unique as there are infinitely many others depending on
an extra free complex parameter $v$ as shown here below,
\begin{equation}
y_{1}=z^{3}v,\qquad y_{2}=xzv,\qquad y_{2}=yv,  \label{c3}
\end{equation}
where $v$ is a homogeneous complex parameter of scaling weight $3$ so that $%
\left( x,y,z,v\right) $ parameterize the $\mathbf{WP}^{3}\left(
3,2,1,3\right) $. Therefore affine $A_{2}$ geometry reads as,
\begin{equation}
\widehat{A}_{2}:y^{2}+x^{3}+z^{6}+\mathrm{\mu }_{0}xyz+v\left( \mathrm{b}%
z^{3}+\mathrm{c}xz+\mathrm{d}y\right) =0.  \label{c4}
\end{equation}
From these relation, one may also write down the vertices and the Mori
charges of the corresponding \ toric polytope; these may be found in \cite
{11}. With the relations (\ref{c1}-\ref{c4}) at hand, we are now ready to
build our first example of complex surface with an indefinite singularity.

\textbf{Over extension of affine }$\mathbf{A}_{2}$\textbf{\ geometry: \qquad
}First of all note that the simplest over extension of $\widehat{A}_{2}$ LKM
algebra is a simply laced indefinite Lie algebra; it is generally denoted as
H$_{3}^{4}$ according to the Classification of Wanglai Lie ( see also
Appendix) and belongs to the so called hyperbolic subset. It has the
following $\mathbb{K}\left( H_{3}^{4}\right) $ Cartan matrix
\begin{equation}
\mathbb{K}\left( H_{3}^{4}\right) =\left(
\begin{array}{cccc}
-2 & 1 & 0 & 0 \\
1 & -2 & 1 & 1 \\
0 & 1 & -2 & 1 \\
0 & 1 & 1 & -2
\end{array}
\right) ,\qquad \mathbb{Q}\left( H_{3}^{4}\right) =\left(
\begin{array}{cccccccc}
-2 & 1 & 0 & 0 & 0 & 0 & 0 & 1 \\
1 & -2 & 1 & 1 & 0 & 0 & 0 & -1 \\
0 & 1 & -2 & 1 & 3 & 1 & 0 & -4 \\
0 & 1 & 1 & -2 & 0 & 1 & 0 & -1
\end{array}
\right)  \label{c5}
\end{equation}
$\mathbb{Q}\left( H_{3}^{4}\right) $ is the matrix of corresponding Mori
vectors to be used later. To get the mirror geometry of a local K3 surface
with $H_{3}^{4}$ singularity, we suppose the three following:

(\textbf{a}) Like for Lie algebra structure where $H_{3}^{4}$ appears as an
over extension of affine $A_{2}$, we consider that hyperbolic $H_{3}^{4}$
geometry is also an extension of affine $A_{2}$ one. As such we conjecture
that the algebraic geometry eq for $H_{3}^{4}$ surface reads as,
\begin{equation}
H_{3}^{4}:y^{2}+x^{3}+\mathrm{\nu }\left( t\right) z^{6}+xyz+\left(
\sum_{i=1}^{4}\mathrm{a}_{i}y_{i}\right) =0,  \label{c6}
\end{equation}
where we have considered an elliptic curve with a varying complex structure.
The $\mathrm{a}_{i}$s moduli describe the complex deformation of $H_{3}^{4}$
singularity of the hyperbolic surface and $y_{i}$s are four gauge invariants
that should be solved in terms of the $x$, $y$, $z$ and $t$ variables.

(\textbf{b}) The relations (\ref{211}) used for affine geometries are also
valid for simply laced indefinite Lie algebra sector. As such we have, for
the present example, the following relations defining the $y_{i}$ gauge
invariants for $H_{3}^{4}$ geometry,
\begin{equation}
\prod_{\alpha =1}^{8}y_{\alpha }^{\mathbb{Q}_{\alpha }^{i}}=1;\qquad
i=1,2,3,4.  \label{c7}
\end{equation}
where the $4\times 8$ rectangular matrix $\mathbb{Q}_{\alpha }^{i}$ define
the four Mori vectors associated with the hyperbolic $H_{3}^{4}$ geometry eq(%
\ref{c5}). The property $\sum_{\alpha }\mathbb{Q}_{\alpha }^{i}=0$ reflects
just the CY condition of this special local K3 surface.

(\textbf{c}) Once the $\mathrm{a}_{i}$ moduli encoding the complex
deformation of hyperbolic $H_{3}^{4}$ surface are taken simultaneously to
zero, the $H_{3}^{4}$ geometry shrinks into eq(\ref{b5}). This means that eq(%
\ref{2110}) should be modified as,
\begin{equation}
y_{n+1}=y^{2},\quad y_{n+2}=x^{3},\quad y_{n+3}=t^{-1}z^{6},\quad
y_{n+4}=xyz,  \label{c9}
\end{equation}
With these tools at hand, one can solve explicitly the remaining four $y_{i}$
gauge invariants in terms of the complex variables $x$, $y$, $z$ and $t$ of $%
\mathbf{WP}^{2}\left( 3,2,1\right) \times \mathbb{C}^{\ast }$. We find,
\begin{equation}
H_{3}^{4}:y^{2}+x^{3}+z^{6}t^{-1}+xyz+\left[ \mathrm{a}z^{6}+\mathrm{b}%
tz^{6}+\mathrm{c}txz^{4}+\mathrm{d}yz^{3}t\right] =0.  \label{c10}
\end{equation}
This is the relation we have been after; it is the mirror of a complex K3
surface with a hyperbolic $H_{3}^{4}$ singularity. CY3 are obtained as usual
by promoting the complex moduli to polynomials depending on an extra
complex\ variable $\zeta $ as $\mathrm{a}_{i}\left( \zeta \right)
=\sum_{j=1}^{n_{i}}\mathrm{a}_{ij}\zeta ^{j}$, where $n_{i}$ stand for the
rank of $U\left( n_{i}\right) $ group symmetries of underlying $4D$ $%
\mathcal{N}=2$ quiver gauge theories that are embedded in type IIA string on
the above CY3 with $H_{3}^{4}$\ singularity. Before concluding, let us give
two more comments: (\textbf{i}) As these kind of unfamiliar CY manifolds
look a little bit unusual, it is interesting to also write down the
solutions for the vertices of the toric polytope associated with the CY3
having a $H_{3}^{4}$\ singularity. They read as,
\begin{eqnarray}
yxz &\longleftrightarrow &V_{8}=\left( 0,0,0\right) ,\quad
y^{2}\longleftrightarrow V_{5}=\left( 0,0,-1\right) ,\quad
x^{3}\longleftrightarrow V_{6}=\left( 0,-1,0\right) ,  \notag \\
t^{-1}z^{6} &\longleftrightarrow &V_{7}=\left( -1,2,3\right) ,\quad
z^{6}\longleftrightarrow V_{1}=\left( 0,2,3\right) ,\quad \
tz^{6}\longleftrightarrow V_{2}=\left( 1,2,3\right) ,  \label{c12} \\
txz^{4} &\longleftrightarrow &V_{4}=\left( 1,1,2\right) ,\quad
yz^{3}t\longleftrightarrow V_{3}=\left( 1,1,1\right) .  \notag
\end{eqnarray}
Using these expressions and eq(\ref{c5}), it is not difficult to check that
these vertices satisfy the basic toric geometry relations namely $%
\sum_{\alpha =0}^{8}q_{\alpha }^{i}=0$ and $\sum_{\alpha =0}^{8}q_{\alpha
}^{i}V_{\alpha }=0$. (\textbf{ii}) The second comment is to discuss the link
between affine $A_{2}$ and hyperbolic $H_{3}^{4}$\ geometries. As noted
before, $H_{3}^{4}$ Lie algebra is just an extension of affine $A_{2}$ and
so one expects that there should be a bridge between the two corresponding
geometries. This is what happens indeed. Starting from the algebraic
geometry eq(\ref{c4}) of affine $A_{2}$, namely $y^{2}+x^{3}+\mathrm{\nu }%
_{0}z^{6}+xyz+$ $\left( \mathrm{b}z^{3}+\mathrm{c}xz+\mathrm{d}y\right) v$,
and performing the following changes
\begin{equation}
\mathrm{\nu }_{0}{\large \longrightarrow }\mathrm{\nu }\left( t\right) =%
\mathrm{\alpha }t^{-1}+\mathrm{a},\qquad v{\large \longrightarrow }v=tz^{3},
\label{c13}
\end{equation}
one gets exactly the hyperbolic $H_{3}^{4}$ mirror geometry of eq(\ref{c10}%
). Here $\mathrm{\alpha }$ and $\mathrm{a}$ are constants.

\textbf{Hyperbolic E}$_{10}$\textbf{\ surface: \qquad }To start recall that
hyperbolic E$_{10}$ is the simplest over extension of affine $E_{8}$. It is
an indefinite KM algebra belonging to the hyperbolic subset, which in Kac
notation, reads as T$_{\left( p,q,r\right) }$ with $\left( p,q,r\right)
=\left( 7,3,2\right) $. Its Cartan matrix $\mathbb{K}\left( E_{10}\right) $
is symmetric and has a\ negative determinant namely $\det \mathbb{K}\left(
E_{10}\right) \mathbb{=-}1$. In 4D $\mathcal{N}=2$ gauge theory embedded in
type IIA string, one may geometric engineer the $E_{10}$ hyperbolic QFT$_{4}$
models and their infrared CFT$_{4}^{\left( -\right) }$ limit by considering
a local CY3 with an $E_{10}$ singularity as outlined in the classification
theorem of section 3. Here we would like to derive $E_{10}$ geometry by
using toric geometry methods and local mirror symmetry. Indeed the
hypothesis of variation of the complex structure of the elliptic curve
allows us to define the hyperbolic $E_{10}$ geometry as,
\begin{equation}
E_{10}:y^{2}+x^{3}+\mathrm{\nu }\left( t\right) z^{6}+xyz+\left(
\sum_{i=1}^{10}\mathrm{a}_{i}y_{i}\right) =0,
\end{equation}
where $\mathrm{a}_{i}$ are complex moduli and where the ten gauge invariant
variables $y_{i}$ are obtained by solving the constraint eqs $\prod_{\alpha
=1}^{14}y_{\alpha }^{\mathbb{Q}_{\alpha }^{i}}=1$. Here $\mathbb{Q}_{\alpha
}^{i}=\mathbb{Q}_{\alpha }^{i}\left( E_{10}\right) $ are the Mori vectors
associated with the hyperbolic $E_{10}$ geometry. $\mathbb{Q}_{\alpha }^{i}$
is a $\left( 10+4\right) \times 10$ rectangular matrix whose $10\times 10$
square block is minus $E_{10}$ Cartan matrix. The solution of the constraint
eqs $\prod_{\alpha =1}^{14}y_{\alpha }^{\mathbb{Q}_{\alpha }^{i}}=1$ may be
obtained without major difficulty as they share features with the product of
the A$_{7}$, A$_{3}$ and A$_{2}$ singularities. Straightforward computations
lead to the following projective exceptional surface,

\begin{equation}
y^{2}+x^{3}+z^{7}t^{-1}+xyz+\mathrm{a}_{0}t^{6}+\mathrm{a}_{1}t^{4}x+\mathrm{%
a}_{2}t^{2}x^{2}+\mathrm{b}_{1}yt^{4}+\sum_{s=1}^{6}\mathrm{c}%
_{s}t^{6-s}z^{s}=0,  \label{h1}
\end{equation}
where $\left( y,x,z,t\right) $ are complex coordinates of WP$_{\left(
3,2,1,1\right) }$. Note that if the $\mathrm{a}_{i}$, $\mathrm{b}_{j}$ and $%
\mathrm{c}_{k}$ complex moduli are simultaneously taken to zero, one ends
with a K3 surface with a hyperbolic $E_{10}$ singularity. Moreover promoting
the $\mathrm{a}_{i}$, $\mathrm{b}_{j}$ and $\mathrm{c}_{k}$ moduli to
polynomials in an extra complex variable $\zeta $ as in eqs(\ref{8}), one
gets a CY3 with complex deformed $E_{10}$ singularity. The degree of these
polynomials define the rank of the gauge quiver group factors, in agreement
with our classification theorem of $\mathcal{N}=2$ CFT$_{4}^{\left( -\right)
}$s. In the end of this study, note that, like for $\widehat{A}_{2}$ and
more generally $\widehat{A}_{r}$, one may here also define a hierarchy of
exceptional geometries; but here these correspond just to the geometries
associated with the so called T$_{\left( p,q,r\right) }$ KM algebra.
Therefore, this kind of algebraic geometric hierarchies are classified by
three positive integers $p$, $q$ and $r$ and the corresponding surfaces are
given by,

\begin{equation}
\left( y^{r}t^{6-3r}+x^{q}t^{6-2q}+z^{p}t^{6-p}+xyz\right) +\mathrm{a}%
_{0}t^{6}+\sum_{s=1}^{p-1}\mathrm{c}_{s}t^{6-s}z^{s}+\sum_{s=1}^{q-1}\mathrm{%
a}_{s}t^{6-2s}x^{s}+\sum_{s=1}^{r-1}\mathrm{b}_{s}y^{r}t^{6-3r}=0,
\label{h3}
\end{equation}
where as before $\left( y,x,z,t\right) $ are in WP$_{\left( 3,2,1,1\right) }$%
. From this relation, one may re-discover known geometries obtained in
earlier literature on 4D $\mathcal{N}=2$ quiver gauge theories. Particular
examples are those associated with finite D$_{r}$, finite E$_{s}$ and affine
$E_{s}$ exceptional geometries. These three classes of geometries correspond
to those T$_{\left( p,q,r\right) }$ algebras with positive determinant of
the Cartan matrices as shown here below,
\begin{equation}
\det \left( \mathbb{K}\left[ T_{\left( p,q,r\right) }\right] \right)
=pq+pr+qr-pqr\geq 0.  \label{dm}
\end{equation}
The remaining subset of T$_{\left( p,q,r\right) }$ algebras with $\det
\left( \mathbb{K}\left[ T_{\left( p,q,r\right) }\right] \right) <0$
corresponds effectively to indefinite geometries; they are described by the
rational number $c=\frac{1}{p}+\frac{1}{q}+\frac{1}{r}<1$. The complex
projective surfaces with $c\geq 1$ are effectively those given by eq(\ref{dm}%
).

\section{Conclusion}

In this paper, we have shown on explicit examples that beta function $%
b_{i}^{\left( q\right) }$ of $\mathcal{N}=2$ quiver gauge theories carries
an extra index $q=1,0,-1$, eq(\ref{t1}). In the infrared limit, these gauge
theories flow to \textit{three different} IR points and so one concludes
that there exist in general three sectors of $\mathcal{N}=2$ CFT$_{4}$s
embedded in type IIA superstring on local CY3s. These sectors are in one to
one with the three classes ( finite, affine and indefinite) of simply laced
KM algebras. Moreover, as these supersymmetric QFT$_{4}$s and their CFT$_{4}$
IR limits are linked with singularities of K3 fibered CY3, we have
conjectured the existence of three kinds of local K3 surfaces classified by
generalized Cartan matrices; one of them has indefinite singularities and
the two others are the well known ones. To illustrate this claim, we have
given two explicit examples namely singular surfaces having hyperbolic H$%
_{3}^{4}$ and E$_{10}$ degeneracies; also known as the over extensions of
affine $A_{2}$ and affine $E_{8}$ respectively. These are given by eqs(\ref
{c10}) and(\ref{h1}). Among our results, we have also found that hyperbolic
geometries may be deduced from the affine category by varying the complex
structure of the elliptic curve on $\mathbb{C}^{\ast }$, see eqs(\ref{c13}).
Extending this idea, we have shown the above hyperbolic singularities are in
fact just leading elements of a hierarchy of a subset of indefinite singular
K3 surfaces obtained by iterative mechanism. For the case of affine $A_{2}$
geometry (\ref{c4}) for instance, one gets upon using eqs(\ref{c13}), the
following surface with deformed hyperbolic $H_{3}^{4}$ singularity,
\begin{equation}
H_{3}^{4}:y^{2}+x^{3}+z^{6}t^{-1}+xyz+\left[ \mathrm{a}z^{6}+\mathrm{b}%
tz^{6}+\mathrm{c}txz^{4}+\mathrm{d}yz^{3}t\right] =0.
\end{equation}
Repeating this procedure once more, one gets the following singular surface $%
y^{2}+x^{3}+z^{6}t^{-2}+xyz+$ $\left[ \mathrm{a}_{-1}z^{6}t^{-1}+\mathrm{a}%
z^{6}+\mathrm{b}tz^{6}+\mathrm{c}txz^{4}+\mathrm{d}yz^{3}t\right] =0$. It is
classified by the following indefinite Lie algebra of minus generalized
Cartan matrix given by,
\begin{equation}
\mathbb{K}\left( H_{3,1}^{4}\right) =\left(
\begin{array}{ccccc}
2 & -1 & 0 & 0 & 0 \\
-1 & 2 & -1 & 0 & 0 \\
0 & -1 & 2 & -1 & -1 \\
0 & 0 & -1 & 2 & -1 \\
0 & 0 & -1 & -1 & 2
\end{array}
\right) ,
\end{equation}
where $H_{3,1}^{4}$ stands for the over extension of $H_{3}^{4}$. Here also,
one can write down the data of this toric manifold as in eqs(\ref{c5},\ref
{c12}). By successive iterations, one may generalize further this result by
constructing the following hierarchy of geometries based on affine $\widehat{%
A}_{2}$,
\begin{equation}
\widehat{A}_{2,k}:y^{2}+x^{3}+z^{6}t^{-k}+xyz+\sum_{s=1}^{k-1}\mathrm{a}%
_{-s}t^{-s}z^{6}+\left[ \mathrm{a}z^{6}+\mathrm{b}tz^{6}+\mathrm{c}txz^{4}+%
\mathrm{d}yz^{3}t\right] =0,\quad k=1,...,  \label{e1}
\end{equation}
where $\widehat{A}_{2,0}$ and $\widehat{A}_{2,1}$ stand respectively for
affine $\widehat{A}_{2}$ and $H_{3}^{4}$\ and where $\widehat{A}_{2,k}$ with
$k>1$ refer for the other hierarchical geometries. Like for T$_{\left(
p,q,r\right) }$ hierarchies we have studied here, see eq(\ref{h3}), the
relations (\ref{e1}) has also poles in $t$. This is a signature of
indefinite geometries.

\begin{acknowledgement}
This work is supported by Protars III, CNRST, Rabat, Morocco. A.Belhaj is
supported by Ministerio de Education cultura y Deporte, grant SB 2002-0036.
\end{acknowledgement}

\section{Appendix: Indefinite Lie algebras}

Indefinite Lie algebras are still a mathematical open subject since their
classification has not yet been achieved. A subset of these indefinite
algebras that is quite well understood includes those known as \textit{%
hyperbolic} Lie algebras \cite{10,12}. According to Wanglai-Li
classification, there are $238$ containing the following special list of
simply laced ones.

\begin{eqnarray}
&&\mathcal{H}_{1}^{4},\quad \mathcal{H}_{2}^{4},\quad \mathcal{H}%
_{3}^{4},\quad \mathcal{H}_{1}^{5},\quad \mathcal{H}_{8}^{5},\quad \mathcal{H%
}_{1}^{6},\quad \mathcal{H}_{5}^{6},\quad \mathcal{H}_{6}^{6},\quad \mathcal{%
H}_{1}^{7},  \notag \\
&&\mathcal{H}_{1}^{7},\quad \mathcal{H}_{1}^{8},\quad \mathcal{H}%
_{4}^{8},\quad \mathcal{H}_{5}^{8},\quad \mathcal{H}_{1}^{9},\quad \mathcal{H%
}_{4}^{9},\quad \mathcal{H}_{5}^{9},\quad \mathcal{H}_{1}^{10},\quad
\mathcal{H}_{4}^{10}.
\end{eqnarray}
For other applications of hyperbolic Lie algebras in string theory; see \cite
{13,14} and refs therein.

\end{document}